\documentclass[aps,amsmath,twocolumn,amssymb,floatfix,showpacs,superscriptaddress,nofootinbib,longbibliography]{revtex4-1}
\usepackage{MnSymbol}
\usepackage{braket}
\usepackage[dvipsnames]{xcolor}
\usepackage{float}
\usepackage{subfigure}
\usepackage{tikz}
\usepackage[colorlinks=true,linktoc=page,linkcolor=red,citecolor=magenta,urlcolor=purple]{hyperref}
\usepackage{amssymb} 
\usepackage{amsfonts}

\mathchardef\mhyphen="2D 

\newcommand{\ie}{{ i.e.,\,\,}}

\newcommand{\mc}{\mathcal}
\newcommand{\mbf}{\mathbf}
\newcommand{\sg}{\sigma}

\newcommand\bea{\begin{eqnarray}}
\newcommand\eea{\end{eqnarray}}
\newcommand\beq{\begin{equation}}  
\newcommand\eeq{\end{equation}}

\usepackage{float}

\newcommand{\non}{\nonumber}  
\usepackage[normalem]{ulem}
\definecolor{lime}{HTML}{A6CE39}
\usepackage{sidecap,tikz}
\usepackage{amsmath} 
\DeclareRobustCommand{\orcidicon}{\hspace{-1.0mm}
	\begin{tikzpicture}
		\draw[lime, fill=lime] (0.0,0.0) 
		circle [radius=0.15] 
		node[white] {{\fontfamily{qag}\selectfont \tiny \,ID}};
		\draw[white, fill=white] (-0.0525,0.095) 
		circle [radius=0.007];
	\end{tikzpicture}
	\hspace{-3.0mm}
}
\foreach \x in {A, ..., Z}{\expandafter\xdef\csname orcid\x\endcsname{\noexpand\href{https://orcid.org/\csname orcidauthor\x\endcsname}{\noexpand\orcidicon}}
}

\AtBeginDocument{%
	\newwrite\bibnotes
	\def\bibnotesext{Notes.bib}
	\immediate\openout\bibnotes=\jobname\bibnotesext
	\immediate\write\bibnotes{@CONTROL{REVTEX41Control}}
	\immediate\write\bibnotes{@CONTROL{%
			apsrev41Control,author="08",editor="1",pages="1",title="1",year="1"}}
	\if@filesw
	\immediate\write\@auxout{\string\citation{apsrev41Control}}%
	\fi
}%

\begin{document}

\title{Identifying Majorana edge and end modes in a Josephson junction of a $p$-wave superconductor with a magnetic barrier}  

\author{Minakshi Subhadarshini\orcidA{}}
\email{minakshi.s@iopb.res.in}
\affiliation{Institute of Physics, Sachivalaya Marg, Bhubaneswar-751005, India}
\affiliation{Homi Bhabha National Institute, Training School Complex, Anushakti Nagar, Mumbai 400094, India}
\author{Amartya Pal\orcidB{}}
\email{amartya.pal@iopb.res.in}
\affiliation{Institute of Physics, Sachivalaya Marg, Bhubaneswar-751005, India}
\affiliation{Homi Bhabha National Institute, Training School Complex, Anushakti Nagar, Mumbai 400094, India}
\author{Pritam Chatterjee\orcidC{}}
\email{pritam.c@iopb.res.in}
\affiliation{Institute of Physics, Sachivalaya Marg, Bhubaneswar-751005, India}
\affiliation{Homi Bhabha National Institute, Training School Complex, Anushakti Nagar, Mumbai 400094, India}
\affiliation{Department of Physics,Graduate School of Science,The University of Tokyo, 7-3-1 Hongo, Bunkyo-ku, Tokyo  113-0033, Japan}
\author{Arijit Saha\orcidD{}}
\email{arijit@iopb.res.in}
\affiliation{Institute of Physics, Sachivalaya Marg, Bhubaneswar-751005, India}
\affiliation{Homi Bhabha National Institute, Training School Complex, Anushakti Nagar, Mumbai 400094, India}


\begin{abstract}

We propose a theoretical model describing a Josephson junction featuring a magnetically textured barrier within two-dimensional (2D) $p$-wave superconductor, considering both $p_x + p_y$ and $p_x + ip_y$ type pairing symmetries. Our study reveals the influence of the magnetic barrier strength and its spatial periodicity on the system's topological properties, in terms of local density of states and Josephson current calculations. Notably, we demonstrate that these parameters regulate the number of Majorana zero modes at the junction in the topological regime. Our setup further allows for the identification of three distinct topological phases, the differentiation between one-dimensional (1D) Majorana edge (either flat/dispersive and arising from intrinsic 2D $p$-wave pairing) and localized Majorana end modes, and an analysis of their hybridization through the Josephson current. In particular, the Josephson current exhibits a discontinuous jump due to the edge modes and pronounced hump in the $p_x + ip_y$ pairing case, directly linked to the hybridized Majorana modes. Moreover, our study opens a possible interesting avenue to distinguish between 1D Majorana edge modes and zero-dimensional end modes via Josephson current signatures.

\end{abstract}

\maketitle

\section{Introduction}
 Majorana fermions are the particles of their own antiparticle that manifest as zero-energy quasiparticles in topological superconductors ~\cite{Kitaev2001,Alicea_2012,Leijnse_2012}. These modes are distinguished by their unique non-Abelian exchange statistics, allowing them for the manipulation of quantum information through braiding operations~\cite{Ivanov2001,SDSarma2008}. Moreover, their intrinsic robustness against local perturbations makes them highly attractive for fault-tolerant topological quantum computation~ \cite{Kitaev2001,Ivanov2001,SDSarma2008,Kitaev2009,Alicea_2012,Leijnse_2012,
beenakker2013search,10.1093/ptep/ptae065,doi:10.1143/JPSJ.81.011013}, as they are less susceptible to decoherence and environmental noise. In current literature, a one-dimensional (1D) $p$-wave superconductors (SCs) provide a fundamental platform for the realization of end localized Majorana zero modes (MZMs), as first demonstrated by Kitaev in his seminal work~\cite{Kitaev2001}. 

In order to engineer such $p$-wave pairing  starting from a conventional $s$-wave SC, several theoretical proposals exists in literature. The first elegant proposal is based on 1D Rashba nanowire, where a semiconducting nanowire with strong Rashba spin-orbit coupling (SOC) in presence of an applied Zeeman field is proximity coupled to a conventional $s$-wave SC~\cite{Oreg2010,DasSharma_2010,Mondal2023}. Very recently, altermagnets~\cite{Smejkal_PRX_1,Smejkal_PRX_2} have been proposed to realize MZMs replacing the external Zeeman field~\cite{Ghorashi2024,Mondal2025}. These setups provide a compelling indirect experimental signatures of MZMs through the zero bias peak in differential conductance measurements~\cite{Oreg2010,DasSharma_2010,Mondal2025,Das2012,Mourik2012}. Beyond Rashba nanowire model, another insightful proposal is utilized in a system consisting of an array of magnetic adatoms deposited on top of a conventional $s$-wave SC in both 1D and 2D domain~ \cite{Felix2013,AliYazdani2013,DanielLoss2013,PascalSimon2013,MFranz2013,Eugene2013,Felix2014,TeemuOjanen2014,MFranz2014,Rajiv2015,Sarma2015,Hoffman2016,Jens2016,Tewari2016,
PascalSimon2017,Simon2017,Theiler2019,Cristian2019,Mashkoori2019,Menard2019,Pradhan2020,Teixeira2020,Alexander2020,Perrin2021,Nicholas2020,Chatterjee2023,chatterjee2023b,Mondal2023,
Jelena2016,Balatsky22016}. In this scenario, the interplay between magnetic impurity spins and the $s$-wave SC gives rise to in-gap Yu-Shiba-Rusinov (YSR) bound states \cite{Felix2013,Yazdani2015,Shiba1968}. The hybridization of YSR states form Shiba band, which effectively mimicks as a $p$-wave SC in the minigap. These engineered heterostructures successfully create the necessary conditions to support topological superconductivity hosting localized MZMs. Later such advancements have significantly propelled the experimental study of Majorana quasiparticles in magnet-SC heterostructures~\cite{Eigler1997,Yazdani1999,Yazdani2015,Wiesendanger2021,Beck2021,Wang2021,Schneider2022,Richard2022,Wiesendanger2022,Yacoby2023,Soldini2023}.

In 2D $p$-wave SCs, the presence of specific pairing symmetries, such as \(p_x + i p_y\) and \(p_x + p_y\), gives rise to 1D Majorana edge modes (MEMs) at their boundary~\cite{Zhang2019, Wang2017, Teemu2015, ZHOU20142576}. For the $p_x + p_y$ pairing symmetry, the system realizes a gapless topological superconductor (TSC) hosting Majorana flat edge modes (MFEMs)~\cite{Zhang2019, Wang2017, Nakosai2013}. Recently, there has been growing interest in gapless TSC hosting MFEMs from both theoretical and experimental perspectives~\cite{Zhang2019, Wang2017, Chatterjee2023_PRBL,Subhadarshini2024,chatterjee2025,bruning2024}. In contrast, for the \(p_x + i p_y\) pairing symmetry~\cite{Teemu2015, ZHOU20142576}, the bulk of the system remains in a gapped topological phase, supporting chiral dispersive edge modes along the 1D edges of the 2D system~\cite{Nakosai2013}. Interestingly, when a 1D magnetic spin texture composed of magnetic adatoms is deposited on the surface of a 2D $p$-wave superconductor, MZMs appear at the ends of the magnetic chain, along with Majorana edge modes at the 2D boundary~\cite{Chatterjee2023}.

Josephson junctions (JJs) are among the most extensively studied systems in condensed matter physics not only because they exhibit intriguing physical phenomena but also due to their promising applications~\cite{Josephson1962,Golubov2004,Shnirman2001,Bergeret2005,Birge2024}. Furthermore, recent advancements in JJs engineered in hybrid superconductor systems have opened new avenues for realizing and manipulating Majorana modes~\cite{Rokhinson2012,Suleymanli_2020,NAKHMEDOV20201353753,Abboud,Kwon2004,FuKane,Fu,YuxanWang,Pal2025}.
In this regard, JJs composed of $p$-wave SCs in presence of a diffusive region and on both sides of a nonmagnetic potential barrier under external magnetic field have been studied~\cite{NAKHMEDOV20201353753,Abboud,PhysRevLett.96.097007}. Additionally, JJs involving unconventional SCs hybridised with topological insulators have also been investigated~\cite{Jacob,Williams}. Very recently, an alternative approach to external magnetic field is utilized by placing magnetic spin textures at the junction between two conventional $s$-wave SCs to induce an effective magnetic field and SOC~\cite{Sardinero}. This effectively generates MZMs localized at the two ends of the junction. However, efffects of such magnetic barrier placed at the junction between 2D $p$-wave SCs are not explored yet. In general, it is very interesting to separate the edge localized Majorana modes and MZMs localized at the end of the magnetic barrier via any suitable response. 

Recent advancements have underscored the potential of 2D platforms in realizing and simulating quantum gates through braiding operations. Notably, simulations of  quantum gates using MZMs localized in magnetic vortices of 2D topological superconductors have been successfully demonstrated~\cite{buss2025braidingmajoranazeromodes}. Furthermore, hybrid architectures combining MEMs and MZMs within vortex cores offer a promising route toward efficient and fault-tolerant quantum gate implementation~\cite{bedow2025majoranaedgemodesquantum}. These developments emphasize the practical relevance of 2D systems for topological quantum computation and provide strong motivation for our current investigation. In particular, a key advantage of 2D models lies in their suitability for braiding unlike 1D systems that rely on complex configurations such as T-junctions to exchange MZMs\,\cite{Alicea_2012}, whereas 2D geometries inherently support adiabatic exchange and direct braiding of MZMs in real space.
\begin{figure}
	\centering
	\includegraphics[width=0.5\textwidth]{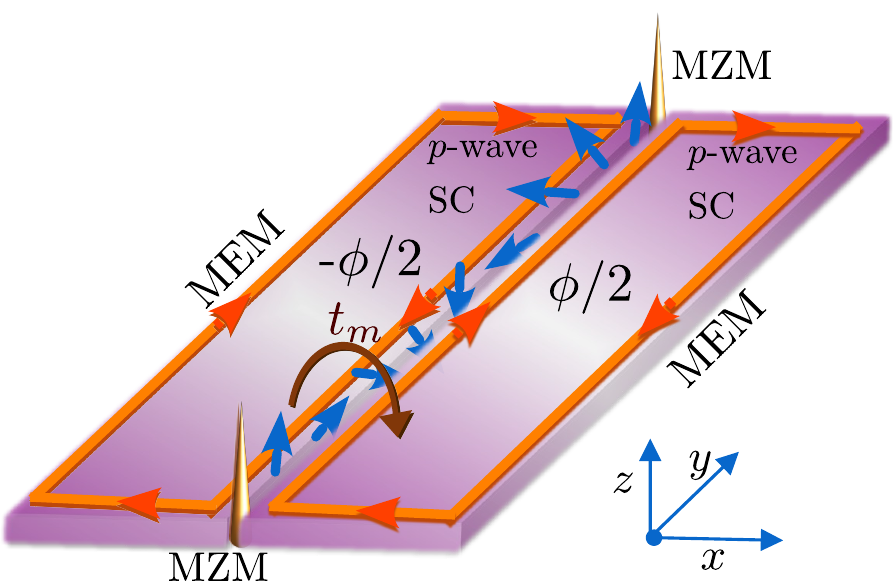}
	\caption{Schematic diagram of our JJ setup is depicted which comprises of two 2D $p$-wave SCs with opposite phase differences of $\pm\phi/2$ in the left and right regions. The junction is characterized by tunneling through a spatially modulated magnetic barrier. MEMs arise in the intrinsic topological $p$-wave SCs, while localized MZMs are generated at two ends of the interfacial magnetic barrier.}
	\label{Fig1}
\end{figure}
Motivated by this fact, in this article, we introduce a theoretical model of a JJ composed of 2D $p$-wave SCs featuring a magnetically textured barrier, as illustrated in Fig.~\ref{Fig1}. We investigate the impact of both $p_x + p_y$ and $p_x + ip_y$ pairing symmetries incorporated in the SC and ask the following intriguing questions: (a) how does the strength and spatial periodicity of the magnetic barrier influence the system's topological properties? (b) how can the Josephson current (JC) possibly identify different Majorana modes present in the system, even if they are hybridized? We intend to answer these questions in this article. We uncover three distinct topological phases revealing the localization of 1D Majorana edge modes at the edges of the SC and MZMs at the ends of the magnetic barrier. Additionally, we analyze the hybridization of these modes via local density of states (LDOS). Furthermore, we also compute the JC associated with these three phases and identify the signatures of each phases. In particular, the Josephson current exhibits a discontinuous jump due to the edge modes and pronounced hump in the $p_x + ip_y$ pairing case, directly linked to the hybridized Majorana modes.

The remainder of the paper is organized as follows. In Sec.~\ref{Sec:II}, we introduce our Hamiltonian of the 2D tight-binding model describing two $p$-wave SCs and a magnetic textured barrier at the junction. In Sec.~\ref{Sec:III}, we discuss the effect of barrier on the band spectrum and the LDOS. Sec.~\ref{Sec:IV} is devoted to the analysis of the impact of the phase difference on LDOS. We discuss the identification of Majorana modes in different topological phases via the JC signal in Sec.~\ref{Sec:V}. We analyze the stability of MZMs and the JC in Sec.~\ref{Sec:VI} by introducing a perturbation at the end of the magnetic barrier coupling to the SCs. Finally, we summarize our findings and conclude in Sec.~\ref{Sec:VII}. The derivation of the effective 1D Hamiltonian and the topological characterization are provided in Appendix~\ref{Appendix-A} and Appendix~\ref{Appendix-B}, respectively. In Appendix~\ref{Appendix-C}, we also discuss the relevance of our results to the corresponding JC behavior of 1D $p$-wave Kitaev chains with a magnetic weak link.

\section{Model}\label{Sec:II}


In this section, we propose a theoretical model of a JJ comprising of two $p$-wave SCs with a superconducting phase difference, $\phi\, (=\phi_R-\phi_L)$, on a square lattice with system dimensions $L_x$ (length) and $L_y$ (width), as illustrated in Fig.~\ref{Fig1}. We introduce the magnetic spin barrier at the junction (i.e. $x=L_x/2)$ along the $y$-direction. Two $p$-wave SCs on either side of the juncton are coupled through a spin dependent hopping amplitude, $t_m$, mimicking the effect of spatially modulated  magnetic spin texture. The tight-binding Hamiltonian for this composite system is written in the real space as,
%
%
%
%
%
\begin{equation}
\mathcal{H}_{JJ} =
\begin{cases} 
\,\,\,\,	\mc{H}_{L}; & 0 \leq x \leq L_x/2,\,\, 0\leq y\leq L_y \\
\,\,\,\,	\mc{H}_{LR}; & x = L_x/2,\,\,  0\leq y\leq L_y \\
\,\,\,\,	\mc{H}_{R}; & L_x/2<x \leq L_x,\,\,  0\leq y\leq L_y\ , \\
\end{cases}
 \label{Eq.BdG_Ham}
\end{equation}

where, $\mc{H}_{L/R}$ represents the left/right $p$-wave SCs on either side of the magnetic barrier and can be written as,~\cite{Zhang2019, Wang2017,Teemu2015,ZHOU20142576}
\begin{eqnarray}
 \mc{H}_{L/R} = -t \!\!\!\!\sum_{<\mbf{r},\mbf{r}^\prime>,\sg}\!\!\!\! c^\dagger_{\mbf{r},\sg}\,c_{\mbf{r}^\prime,\sg}\,\!\!\! &+&\! \!\!\! \sum_{<\mbf{r},\mbf{r}^\prime>,\sg} \!\!\! \Delta_{\mbf{r},\mbf{r}^\prime} e^{i\phi_{L/R}}\, c^{\dagger}_{\mbf{r},\sg} c^{\dagger}_{\mbf{r}^\prime,\sg}   + \rm{h.c} \non \\&-& \mu \sum_{\mbf{r},\sg}  c^\dagger_{\mbf{r},\sg}\,c_{\mbf{r},\sg}\ ,
\end{eqnarray}
where, $c_{\mbf{r},\sigma}$ and $c_{\mbf{r},\sigma}^\dagger$ denote the annihilation and creation operators for electrons with spin $\sigma$ at site $\mbf{r}=(x,y)$. Model parameters $t$ and $\mu$ denote the hopping amplitude and chemical potential of the two $p$-wave SCs respectively. The superconducting phase difference between the right and left SCs is $\phi=\phi_R-\phi_L$. Importantly, the superconducting pairing amplitude, $\Delta_{\mbf{r},\mbf{r}^\prime}$ is different for $(p_x+p_y)$- and $(p_x+ip_y)$ pairing symmetry and given as,~\cite{Zhang2019, Wang2017,Teemu2015,ZHOU20142576}
\begin{equation*}
	\Delta_{\mbf{r},\mbf{r}^\prime}=\!\!(\Delta_{x,x+a},\Delta_{y,y+a}) =
	\begin{cases} 
		(\Delta_0,\Delta_{0}) \mathrm{\,\, for \,\,} (p_x + p_y)  \mathrm{\,\, pairing} \\
		(\Delta_0,i\Delta_{0}) \mathrm{\,\, for \,\,} (p_x + ip_y)  \mathrm{\,\, pairing} \\
	\end{cases}
\end{equation*}
Here, $\Delta_{0}$ denotes the bulk superconducting pairing amplitude of the considered $p$-wave SCs. Note that, $(p_x+p_y)$ and $(p_x+ip_y)$ pairing exhibit distinct topological band properties as we discuss later.

Magnetic texture coupled tunneling Hamiltonian ($H_{LR}$) at the junction between left and right $p$-wave SCs can be written as:~\cite{Sardinero}
\begin{equation}
	\mc{H}_{LR} = \sum_{y,\sigma,\sigma'} c_{L_x/2,y,\sigma}^\dagger \tilde{t}_{y}^{\sigma\sigma'} c_{L_x/2+a,y,\sigma'} + \text{h.c.}\ ,
	\label{Eq. Coupling_Ham_H_LR}
\end{equation}
where, \( \tilde{t}_{y}^{\sigma\sigma'} \) represents the tunneling matrix element of the following \( 2 \times 2 \) matrix in spin space, incorporating the magnetic texture barrier and is given by,
\begin{equation}
	\tilde{t}_y = t_0 \sigma_0 + {\boldsymbol{t}}_y \cdot \boldsymbol{\sigma}\ ,
	\label{Eq.Magnetic_barrier1}
\end{equation}
where $t_0$ denotes the amplitude of the spin-conserving term, and ${\boldsymbol{t}}_y $ represents the spin-dependent components with $\boldsymbol{\sigma} = (\sigma_x, \sigma_y, \sigma_z)$ being the Pauli matrices in spin space.

The spin-dependent component of magnetic barrier is expressed as:
\begin{equation}
	\boldsymbol{t}_y = -t_m \left( \sin \theta_y \cos \phi_y \sigma_x , \, \sin \theta_y \sin \phi_y \sigma_y , \,\cos \theta_y \sigma_z \right)\ ,
\end{equation}
where, $\phi_y$ and $\theta_y$ denote the azimuthal angle and the polar angle between two spins respectively along the $y$ direction. Here, $t_m$ is the magnitude of the spin-dependent hopping. 

Further, we assume that the rotation of the spin is confined to the \( xz \)-plane, where \( \phi_y = 0 \). Therefore, \( \boldsymbol{t}_y \) takes the following form,
\begin{equation}
	\boldsymbol{t}_y = -t_m \left[ \sin\left(\frac{2\pi y}{\xi_m}\right) \sigma_x ,\,0,\,\cos\left(\frac{2\pi y}{\xi_m}\right) \sigma_z \right]\ .
\end{equation}

where, $\xi_m$ is called spatial modulation vector which determines the spatial periodicity of the spin texture along the $y$-direction.

\section{Effect of Magnetic Barrier}\label{Sec:III}
In this section, we investigate the impact of the magnetic barrier formed by a spatially modulated spin texture on the topological properties of the JJ comprised of 2D $p$-wave SCs. We first set the superconducting phase difference, $\phi=0$, to understand the sole effect of the magnetic barrier whereas the effect of phase bias will be discussed in the next section. We analyze the topological properties of the JJ by computing the energy eigenvalue spectra and site-resolved normalized zero energy LDOS. For this purpose, we first numerically diagonalizing the Hamiltonian [Eq.\eqref{Eq.BdG_Ham}] in a 2D finite size system with dimension ($L_x,L_y$) employing the open boundary condition (OBC) and obtain the site resolved LDOS, $N(\mbf{r},\omega)$ from the imaginary part of the retarded Green's function as, 
%
\begin{equation}
	N(\mbf{r},\omega) = -\frac{1}{\pi} \mathrm{Im[Tr}\,\mc{G(\mbf{r},\mbf{r},\omega)}]\ ,
	\label{Eqn:LDOS}
\end{equation} 
where, $\mc{G}(\mbf{r},\mbf{r^\prime},\omega)$ is the retarded Green's function and obtain using Eq.~\eqref{Eq.BdG_Ham} as, 
\begin{equation}
	\mc{G}(\mbf{r},\mbf{r^\prime},\omega) = [(\omega + i\delta)\mbf{I} - \mc{H}_{\rm{JJ}}(\mbf{r},\mbf{r^\prime})]^{-1}\ .
\end{equation}

Here, we mainly focus on the zero energy LDOS i.e. $N(\mbf{r},\omega=0)$  in the $x-y$ plane to characterize and understand the localization profile of MEMs as well as MZMs.
We set $t=1,~\Delta_{0}=0.5t,~t_0=1$, and $L_x=L_y=50$ lattice sites thoughout the paper. Other model parameters values are mentioned explicitly. The results of this analysis are presented and 
discussed below.

\begin{figure}
	\centering
	\subfigure{\includegraphics[width=0.5\textwidth]{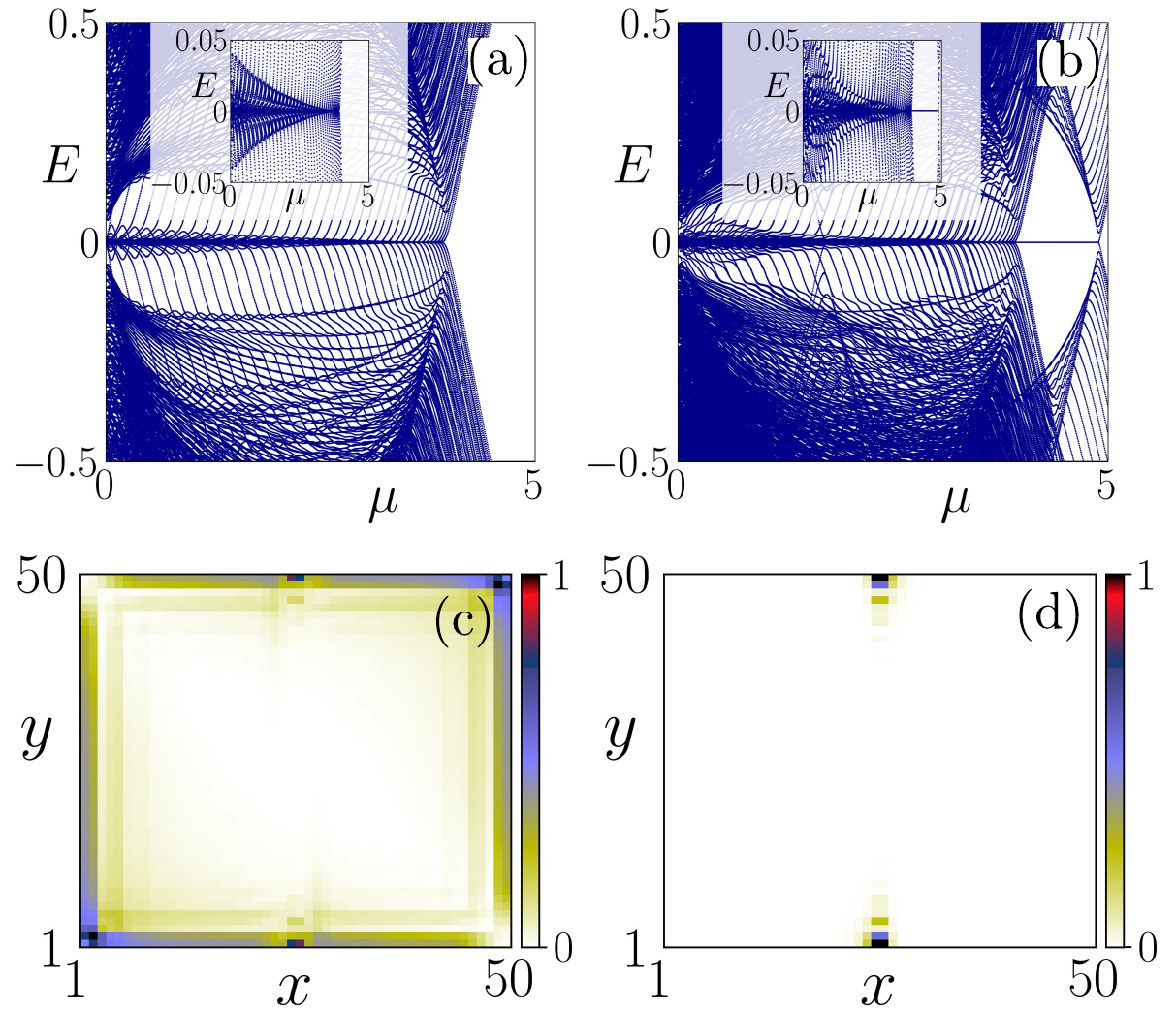}}
	\caption{We illustrate the energy eigenvalue spectrum in the absence ($t_m=0$) and presence ($t_m\ne0$) of a magnetic barrier for a $p_x+p_y$ type SC in panels (a) and (b) respectively. The insets of panel (a) and (b) exhibits the zoomed version of eigenvalue spectrum near zero-energy for better clarity. Panels (c) and (d) depict the LDOS for $\mu = 3t$ and $\mu = 4.2t$ under the influence of the magnetic barrier ($t_m\ne0$). For panels (b), (c), and (d), the barrier strength is chosen as $t_m = 2t$. The calculations are performed using OBC and on a finite size $50 \times 50$ square lattice. The other model parameters are chosen as $\Delta_0 = 0.5t$, $\xi_m = 5$, $t_0 = t$, and $t = 1$.
	}
	\label{Fig2}
\end{figure}

\begin{figure}
	\centering
	\subfigure{\includegraphics[width=0.5\textwidth]{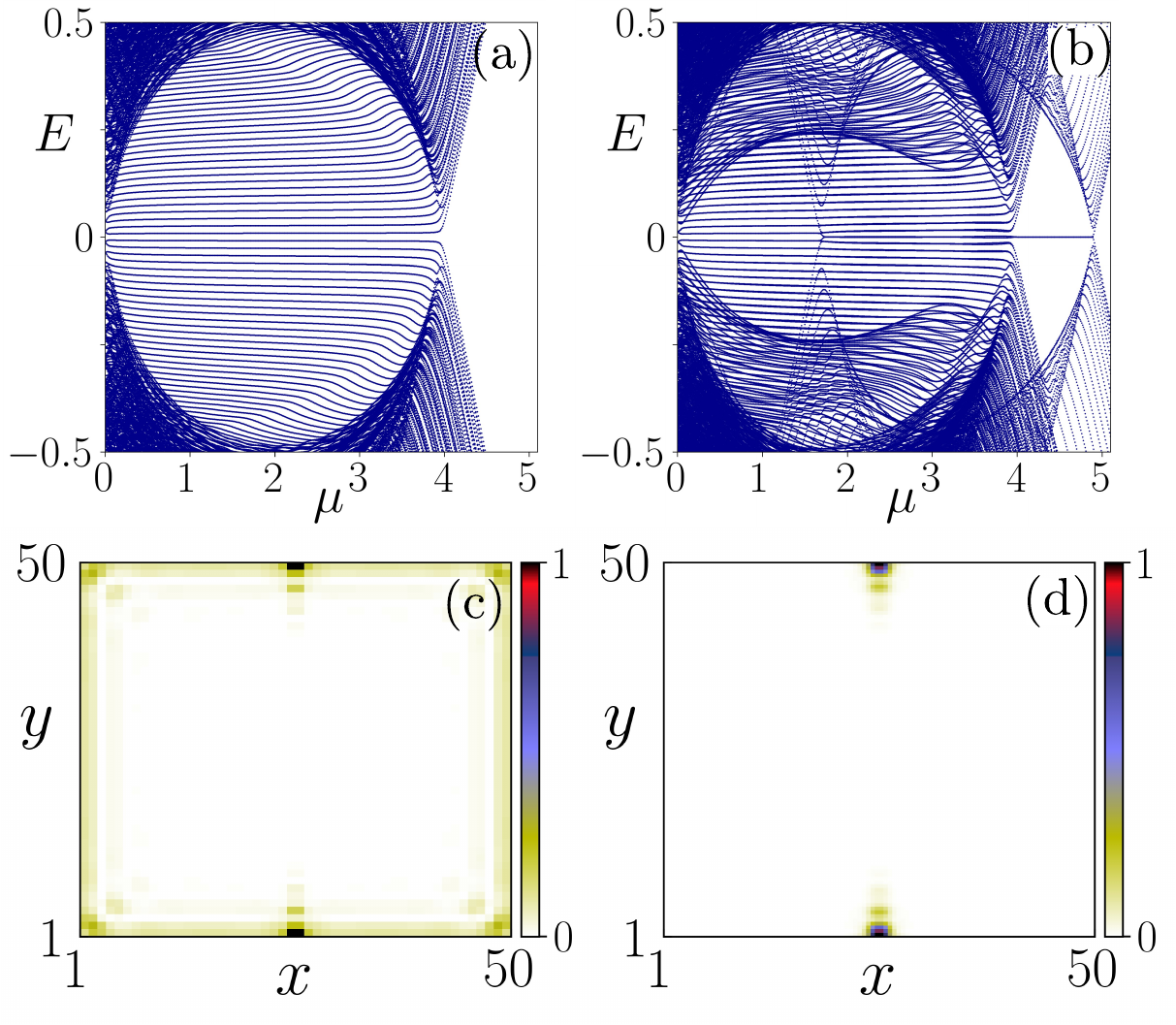}}
	\caption{In case of $p_x + ip_y$ type SC, panel (a) and (b) represent the energy eigenvalue spectrum without and with the presence of magnetic barrier, respectively. Panels (c) and (d) highlight the LDOS distribution for $\mu = 3t$ and $\mu = 4.2t$ under the effect of the magnetic barrier. The barrier strength is uniformly chosen as $t_m = 2t$ for panels (b), (c), and (d). The numerical simulations are performed on a $50 \times 50$ square lattice, with same model parameters considered for Fig.~\ref{Fig2}.}
	\label{Fig3}
\end{figure}

\subsection{For $p_x+p_y$ type pairing}
A ($p_x \!+\! p_y$)-wave SC harbors intrinsic gapless topological superconductivity hosting MFEMs at zero energy~\cite{Zhang2019, Wang2017}. In Fig.~\ref{Fig2}(a), we plot the energy eigenvalue spectra of Hamiltonian [Eq.\eqref{Eq.BdG_Ham}] as a function of chemical potential, $\mu$. We clearly observe the existence of zero energy MFEMs for $\mu<4t$ showcasing the topology of 2D 
($p_x+p_y$)-wave SC. For $\mu>4t$, the bulk of the system becomes gapped and MFEMs also disappear from the system. In absence of the magnetic barrier ($t_m=0$) only the spin-conserving hopping parameter, $t_0$, in Eq.~\eqref{Eq.Magnetic_barrier1} contributes. In this case, the two SCs behave as a single ($p_x + p_y$)-wave SC, and consequently, zero-energy modes localize at the edges, corresponding to MFEMs which is schematically illustrated in Fig.~\ref{Fig1}. Now, when the strength of the magnetic barrier is turned on, two other topological phases emerge in the system. 
In Fig.~\ref{Fig2}(b), we illustrate the eigenvalues spectra for $t_m=2t,~\xi_m=5$ as a funciton of $\mu$. First. we observe that for $\mu_{c1}<\mu<4t$, the system hosts both MZMs and MFEMs. Note, for $t_m=2t$, we find $\mu_{c1}\simeq 1.64 t$ which, in general, will depend on the values of $t_m$ and $\xi_m$. Presence of such mixed topological phase, is further confirmed by computing the LDOS, $N(\mbf{r},E=0)$. We plot the $N(\mbf{r},E=0)$ in the $x-y$ plane for $\mu=3t$ in Fig.~\ref{Fig2}(c) and observe that the MZMs are localized at the ends of the magnetic barrier whereas MFEMs are localized along the edges of the 2D domain. Note that, the junction MFEMs originating from each of the isolated $p_x+p_y$-type SC are now gapped out due to the coupling Hamiltonian, $\mc{H}_{LR}$ [see Eq.\,\eqref{Eq. Coupling_Ham_H_LR}]. Furthermore, from Fig.~\ref{Fig2}(b) we observe the presence of only MZMs beyond $\mu=4t$ confined solely to the end of the magnetic barrier, while the MFEMs of the $p$-wave SC are completely gapped out. This is further assured from LDOS calculation as shown in Fig.~\ref{Fig2}(d) where we plot the $N(\mbf{r},E=0)$ for $\mu=4.2t$. These observations suggests the appearance of two new topological phases and more importantly extension of topological regime beyond $\mu=4t$ compared to bare $p$-wave SC. We enlist these three topological phases below,

\begin{itemize}
	\item[(i)] MFEMs localized at the 2D boundary for $\mu < \mu_{c1}$ belonging to the parent $p$-wave SC, 
	 \item[(ii)] Hybrid topological phase where both MFEMs and MZMs coexist for $\mu_{c1} < \mu < 4t$, and
	\item[(ii)] Isolated MZMs localized at the end of the magnetic barrier for $\mu > 4t$.
\end{itemize}

For the purpose of topological characterization of these phases hosting different modes, we derive an effective 1D Hamiltonian by setting \( L_x = 2 \) at the junction and considering the limit \( L_y \rightarrow \infty \) (see Appendix~\ref{Appendix-A} for details), thereby reducing the full 2D system to two superconducting chains coupled along the $y$-direction via the magnetic barrier. For the resulting effective Hamiltonian \( H_{\text{eff}} \) (see Appendix~\ref{Appendix-B2}), we compute the topological invariant, the winding number \cite{Chiu2016RMP,Mondal2023b} in the \( t_m \)-\( \xi_m \) parameter space. This winding number characterizes the presence of MZMs at the end points of the junction barrier. The invariant \( W \) takes values 0, 1, or 2, corresponding respectively to: (i) a trivial phase with no MZMs, (ii) the presence of two MZMs (one at each end of the junction), and (iii) four MZMs (two at each end) as discussed in Appendix~\ref{Appendix-A}.

\begin{figure}
	\centering
	\subfigure{\includegraphics[width=0.5\textwidth]{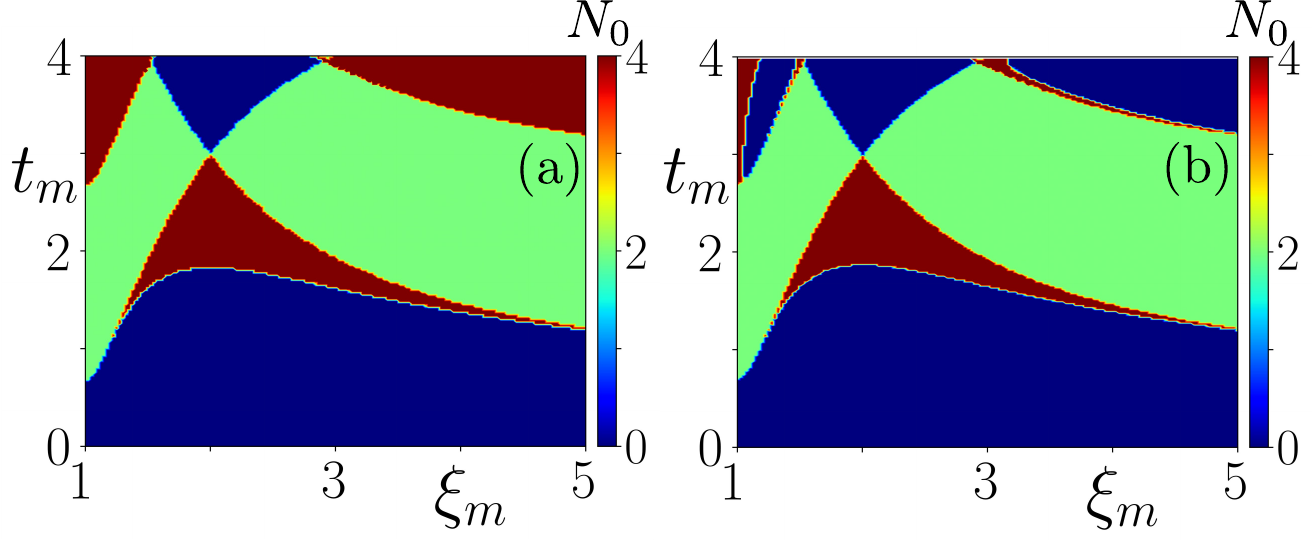}}
	\caption{In panels (a) and (b), we show the variation in the number of MZMs within the $t_m - \xi_m$ plane considering the $p_x + p_y$ and $p_x + ip_y$ type SCs, respectively. The results are obtained for the model  parameters $\mu = 4.2t$, $t = 1$, $t_0 = t$, and $\Delta_0 = 0.5t$. The numerical computation is performed on a $50 \times 50$ square lattice.
	}
	\label{Fig4}
\end{figure}



\subsection{For $p_x+ip_y$ type pairing}

In contrast to $(p_x+p_y)$-wave SC, bulk $(p_x+ip_y)$-wave SC features a gapped spectrum and hosts gapless chiral dispersive MEMs~\cite{Teemu2015,ZHOU20142576}. Similar to the $(p_x + p_y)$-wave SC, we analyze the appearence of various topologial phases using eigenvalue spectrum and LDOS calculation. First, in absence of any magnetic barrier ($t_m=0$), the topological regime is presented for $\mu<4t$ as demonstrated in Fig.~\ref{Fig3}(a). However, when the spin-dependent tunneling barrier ($t_m \neq 0$) is introduced, the topological region extends beyond $\mu > 4t$. Furthermore, three distinct topological phases emerge even in this scenario: The dispersive MEMs ($\mu<\mu_{c2}$), MZMs hybridized with the MEMs ($\mu_{c2}<\mu<4t$), and only MZMs ($\mu>4t$). Similar to the earlier case, the value of $\mu_{c2}$ depends on $t_m$ and $\xi_m$ and for $t_m=2t,~\xi_m=5$,~$\mu_{c2}\simeq 1.76t$. These topological phase transitions are associated with the bulk band closings at two topological phase transition points, as observed in Fig.~\ref{Fig3}(b).

The LDOS, computed for these emergent topological phases, is shown in Figs.~\ref{Fig3}(c), (d). Here, Fig.~\ref{Fig3}(c) reveals the coexistence of both MEMs and MZMs, localized at the edges of the 
2D domain and ends of the magnetic barrier simultaneously. Whereas from Fig.~\ref{Fig3}(d) we conclude that the previously existing MEMs are gapped out, while the MZMs are localized at the ends of the magnetic barrier. 

In the region where only MZMs exist ($\mu>4t$), the number of modes can be tuned due to their clear separation from the bulk states. By tuning the parameters $t_m$  and $\xi_m$, the number of Majorana modes in the topological regime can be precisely controlled. This tunability enables a transition from 2 to 4 MZMs in case of both $p_x + p_y$ and $p_x + ip_y$ type SCs, as shown in Figs.~\ref{Fig4}(a) and (b), respectively. The reason for appearance of multiple MZMs can be attributed to the fact that tuning of $t_m$ and $\xi_m$ can generate higher order hopping resulting in multiple MZMs. However, the phase diagram exhibits slight differences between the two cases, reflecting different characteristics of each $p$-wave superconducting system.

In this case, $p_x + i p_y$ pairing breaks time reversal symmetry and we compute the Chern number using the Fukui method~\cite{Fukui_2005} to topologically characterize the phase. Specifically, we consider the case where only MEMs are present, with no spin-dependent tunneling present (i.e., \( t_m = 0 \)). 
As discussed in Appendix~\ref{Appendix-B1}, the Chern number takes the value \( C = 1 \) in the region \( 0 < \mu < 4t \), which corresponds to the edge localized dispersive MEMs depicted in Fig.~\ref{Fig3}(a). 

\begin{figure}
	\centering
	\subfigure{\includegraphics[width=0.5\textwidth]{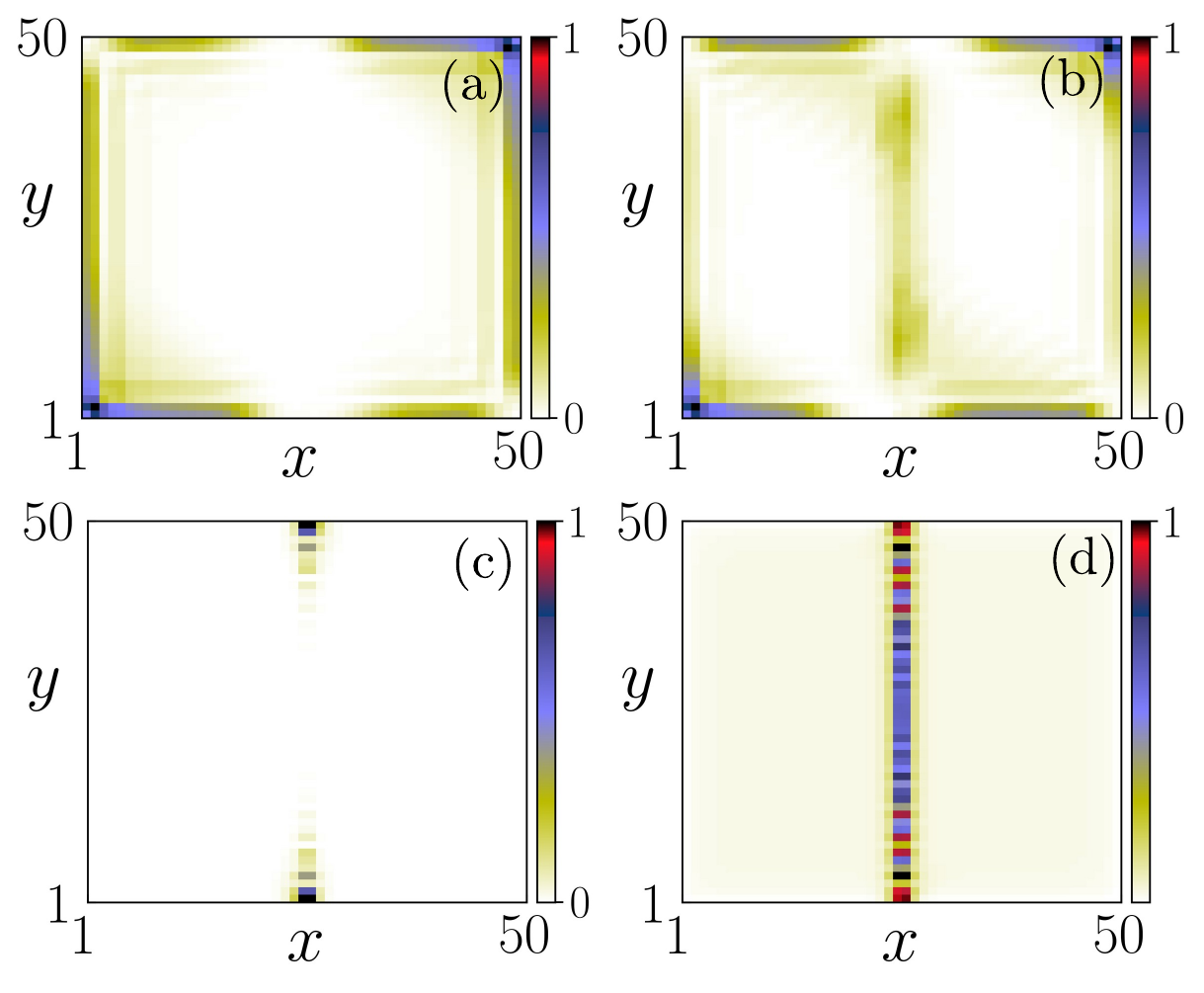}}
	\caption{LDOS spectra at \(E = 0\) is shown for MEMs at \(\mu = 3t\) and MZMs at \(\mu = 4.2t\), applying two different phase differences in case of \(p_x + p_y\) SC. Panels (a) and (c) correspond to LDOS for \(\phi = \pi/2\), while panels (b) and (d) represent the same for \(\phi = \pi\). The model parameters used are \(t_m = 2t\), \(t = 1\), and \(\Delta_0 = 0.5t\). We consider a finite system size of $50\times50$ square lattice.}
	\label{Fig5}
\end{figure}
%
\section{Effect of Phase Difference on LDOS}\label{Sec:IV}
In this section, we investigate the effect of the phase difference, $\phi$, between the left and right SCs on the zero energy LDOS, $N(\mbf{r},0)$, [see  Eq.~\eqref{Eqn:LDOS}], considering both types of $p$-wave pairings. The superconducting phase of the left and right SCs are chosen to be, $\phi_R=-\phi_L=\phi/2$ which maintains a phase difference of $\phi$.  Particularly, we analyze the effect of phase difference on the emergent topological phases as discussed in the previous section. These results provide insights into the behavior of the system in the presence of different pairing symmetries and phase configurations.

\subsection{For $p_x+p_y$ type pairing}
First, we consider the $p_x+p_y$-type pairing and illustrate the behaviour of $N(\mbf{r},0)$ in the $x-y$ plane as depicted in Fig.\,\ref{Fig5}. In Fig.\,\ref{Fig5}(a) and \ref{Fig5}(b), we plot the $N(\mbf{r},0)$ in the hybrid topological phase for  $\phi=\pi/2$ and $\phi=\pi$, respectively fixing $\mu=3t$, $t_m=2t,$ and $\xi_m=5$. Interestingly, in the hybrid topological phase the MZMs that are localized at the ends of the magnetic barrier, are now gapped out due to the applied phase bias as shown in Fig.\,\ref{Fig5}(a). This suggests that tuning the phase difference, the system moves from hybrid topological phase to the phase with only MFEMs. Similar to $\phi=0$ [Fig.\,\ref{Fig2}(c)], at $\phi=\pi/2$, the junction MFEMs still remains gapped. Then, in Fig.\,\ref{Fig5}(b) for $\phi=\pi$, we also observe the absence of MZMs at the ends of magnetic barrier. However, now a bulk channel appears near the junction which is localized along the $y$-direction. Making analogy with 1D $p$-wave SCs, we find the energy splitting of junction localized Majorana modes, $\Delta E_{jun}\sim \cos (\phi/2)$~\cite{Alicea_2012}. Therefore, for $\phi=\pi$, the junction MFEMs becomes degenerate at zero energy and appears to be localized along the junction irrespective of the barrier strength, as can be seen in the LDOS profile of Fig.\,\ref{Fig5}(b). In contrast to the hybrid topological phase, in Fig.~\ref{Fig5}(c), when only MZMs exist, the LDOS exhibits no significant difference compared to \(\phi = 0\) case [see Fig. \ref{Fig2} (d)]. Moreover, a channel forms through the magnetic barrier as the band gap closes at $\phi=\pi$ as shown in Fig.\,\ref{Fig5}(d). This effectively connects the localized end modes through the bulk. This feature is observed in both Figs.~\ref{Fig5}(b) and \ref{Fig5}(d).  Additionally, in Fig.~\ref{Fig5}(c) and (d), the MFEMs are completely gapped out.
%
%
%

\begin{figure}
	\centering
	\subfigure{\includegraphics[width=0.5\textwidth]{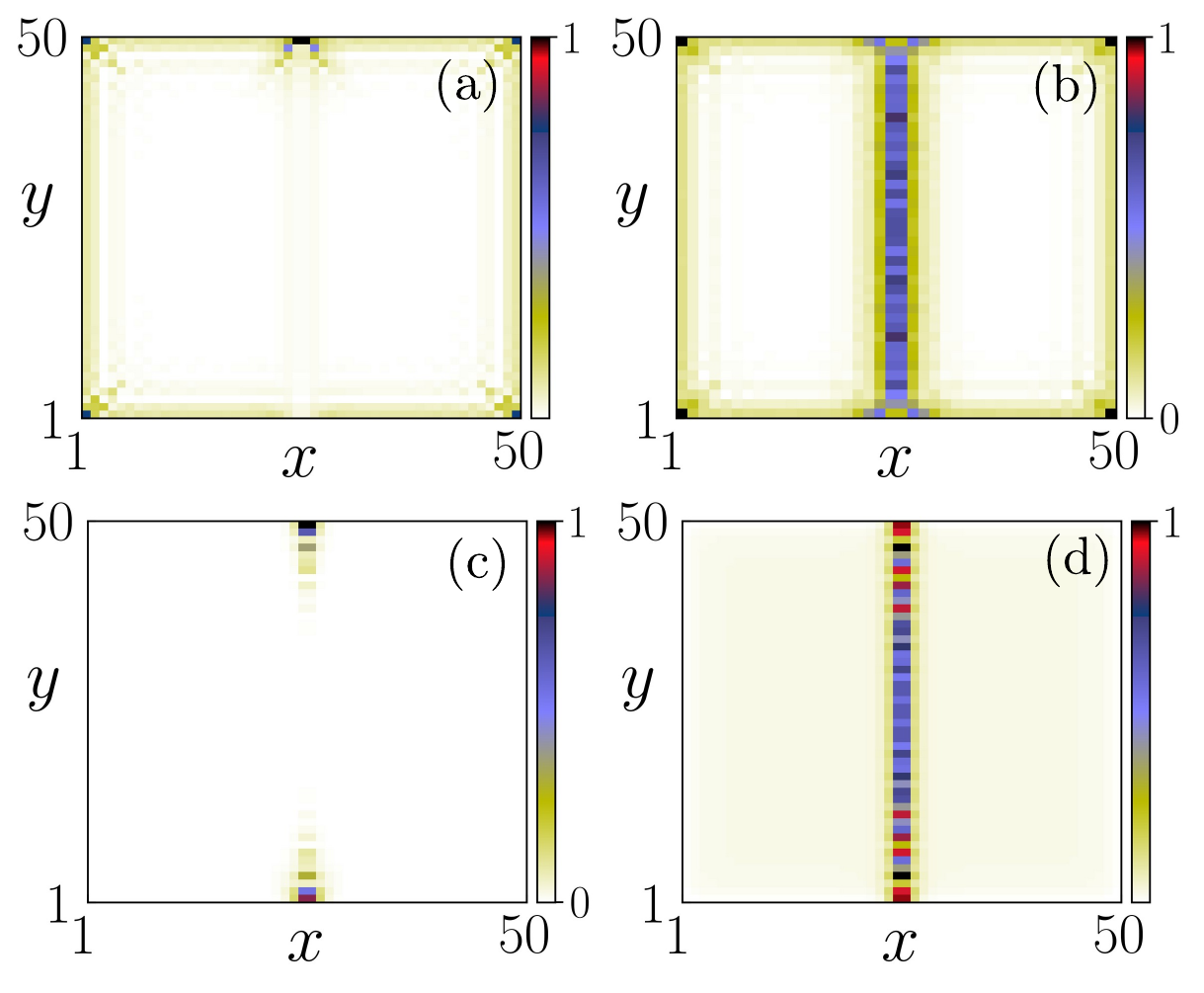}}
	\caption{LDOS distribution at \(E = 0\) is depicted for \(p_x + ip_y\) SC, showcasing the behavior  of MEMs at \(\mu = t\) and MZM at \(\mu = 4.2t\) under the application of two phase differences. Panels (a) and (c) correspond to a phase difference of \(\phi = \pi/2\), while panels (b) and (d) display the results for \(\phi = \pi\). The numerical calculations are performed considering a $50\times50$ square lattice and for $t_m=2t$. The other model parameters are chosen to be the same as mentioned in Fig.~\ref{Fig5}.}
	\label{Fig6}
\end{figure}
%
\subsection{For $p_x+ip_y$ type pairing}
%
Here, we discuss the effect of phase bias on LDOS profile considering $p_x+ i p_y$-type SC and illustrate our findings in Fig.\,\ref{Fig6}. In Fig.\,\ref{Fig6}(a) and \ref{Fig6}(b), we show zero energy LDOS in the $x-y$ plane for $\phi=\pi/2$ and $\phi=\pi$ respectively with $\mu=t, t_m=2t,$ and $\xi_m=5$. In Fig.\,\ref{Fig6}(a), we observe the presence of MZMs at the top end of the magnetic barrier along with the MEMs localized along the edges of the system. Interestingly, for the same values of $\mu,t_m$, and $\xi_m$ the system hosts only MEMs in absence of any phase difference \ie $\phi=0$. Therefore, for $p_x + i p_y$-type pairing, introducing a phase bias drives the system into hyrid topological phase where both MEMs and MZMs coexist. Next, for $\phi=\pi$, the system hosts MEMs at 
the junction along the $y$-direction as depicted in Fig.\ref{Fig6}(b). This has the similar origin as the $p_x+p_y$-type pairing where the junction MEMs becomes gapless for $\phi=\pi$. At \(\phi = \pi/2\), Fig.~\ref{Fig6}(c) exhibits similar behavior to the \(p_x + p_y\)-type pairing [see Fig \ref{Fig5}(c)] where the topological region corresponds to only MZMs. While at $\phi=\pi$, the intermediate channel coexists with MZMs in the absence of edge modes [see Fig. \ref{Fig6}(d)]. The reason is at $\phi=\pi$, the spectrum becomes gapless and this channel appears due to the formation of junction localized bulk states.
%

\section{Analysis of Majorana Modes through Josephson Current}\label{Sec:V}
In order to analyze the signatures of MEMs, MZMs, and the appearance of the  hybrid topological phase, we compute the JC across the junction. The zero-temperature JC can be computed using the following relation~\cite{BeenakkerJosephson}:
\begin{equation}
I(\phi) = \frac{2 e}{\hbar}\sum_{\epsilon_i < 0} \partial_\phi \epsilon_i(\phi)\ ,
\end{equation}
where the summation is performed over all the occupied energy eigenstates. The energy eigenvalues, $\epsilon_i$, are obtained by performing exact diagonalization of the Hamiltonain described in 
Eq.\,\eqref{Eq.BdG_Ham}. We investigate the dependence of the JC on key parameters, such as the chemical potential (\(\mu\)) and the barrier strength (\(t_m\)), and the corresponding results are discussed in detail below.

\subsection{Effect of Chemical Potential}\label{Sec:VA}
\begin{figure}
	\centering
	\subfigure{\includegraphics[width=0.5\textwidth]{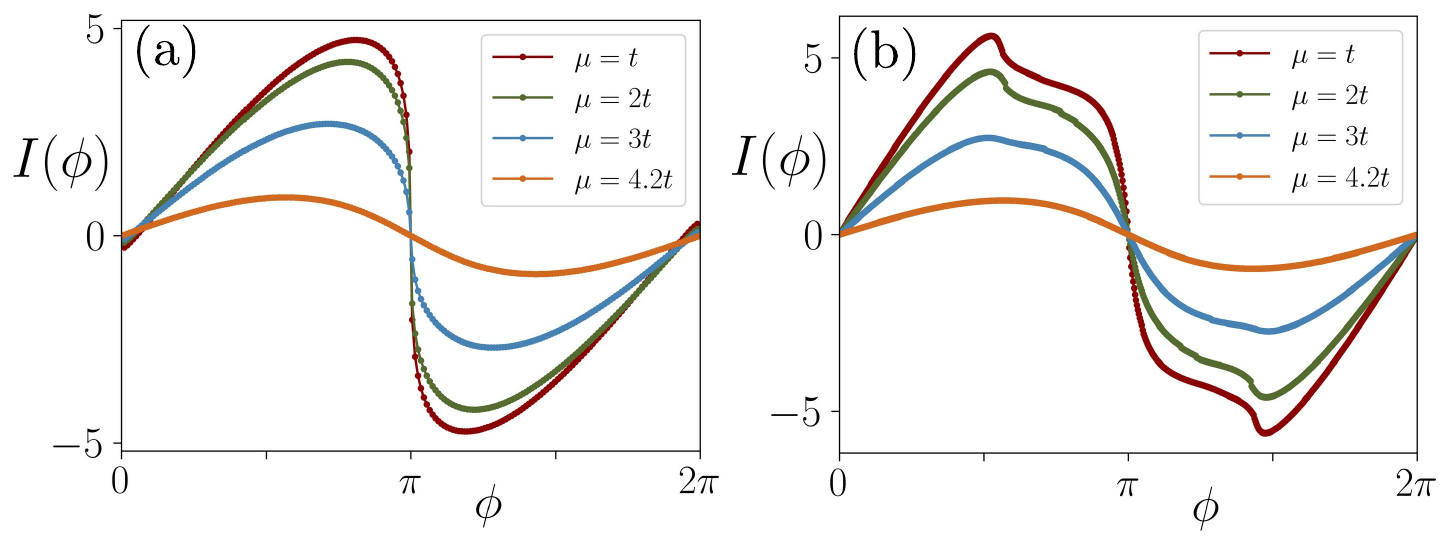}}
	\caption{In panel (a), we depict the variation of the JC as a function of $\phi$ for different values of the chemical potential (\(\mu\)) for a \(p_x + p_y\) SC, while in panel (b) we illustrate the corresponding behavior for a \(p_x + ip_y\) SC. The model parameters used for these computations are \(\Delta_{0} = 0.5t\), \(t_{0} = t\), \(t = 1\), and \(t_m = 2t\). We consider a finite size system 
	of $50\times50$ square lattice.}
	\label{Fig7}
\end{figure}

We first discuss the effect of $\mu$ on JC profile considering $p_x + p_y$-type pairing for a fixed value of $(t_m,\xi_m) = (2t,5)$ and showcase the behaviour of $I(\phi)$ as a function of phase bias, $\phi$, in Fig.\,\ref{Fig7}(a). For $\mu=t$ i.e. when the system only hosts MEMs, we observe a sharp jump at $\phi=\pi$ and the maximum value of $I(\phi)$ appears close to $\phi=\pi$. This sharp jump emerges due to change of parity of the superconducting ground state and leads to the $4\pi$-periodic Josephson effect, attributed to the topological nature of $p$-wave SC~\cite{Alicea_2012}. 
By varying $\mu$ from $t$ to $4.2t$, first the system makes transitions to the hybrid topological phase characterized by the presence of both MEMs and MZMs, and then to one dominated exclusively 
by localized MZMs. The sharp jump observed for $\mu=2t$, becomes reduced in the hybrid topological phase and the maximum value of the current gets shifted towards $\phi=\pi/2$. This sharp jump gradually diminishes as the chemical potential increases, reflecting the progression toward the phase dominated by solely end modes. At \(\mu = 4.2t\), where only end modes are present, this sharp transition around \(\phi = \pi\) becomes vanishingly small and the $I(\phi)$ exhibits $2\pi$-perodic behaviour as the MEMs are completely gapped out. 

On the other hand, for the $p_x + ip_y$-type SC, the behavior of $I(\phi)$ is notably different compared to the $p_x + p_y$-type pairing as shown in Fig.~\ref{Fig7}(b). For $\mu=t,2t,$ and $3t$, the system harbors the hybrid topological phase as elaborately discussed in the previous section [see Fig.\,\ref{Fig6}(a)]. In this case, a hump in $I(\phi)$ is observed near $\phi=\pi/2$ indicative of the hybridized modes unique to this pairing symmetry. Note that, this feature is absent in the \(p_x + p_y\) case. As the system transits towards the phase dominated by MZMs (\(\mu > 4t\)), the hump gradually diminishes and the system shows a $2\pi$-periodic Josephson effect. At \(\mu = 4.2t\), the behavior of the \(p_x + ip_y\) type SC closely resembles that of the \(p_x + p_y\) type SC, as both 
are then governed by MZMs. In that scenario, JC is carried by the bulk Cooper pairs only for both $p_x+ip_y$ and $p_x+p_y$ case and end modes don't contribute to the JC.

\subsection{Effect of Magnetic Barrier Strength}\label{Sec:VB}
\begin{figure}
	\centering
	\subfigure{\includegraphics[width=0.5\textwidth]{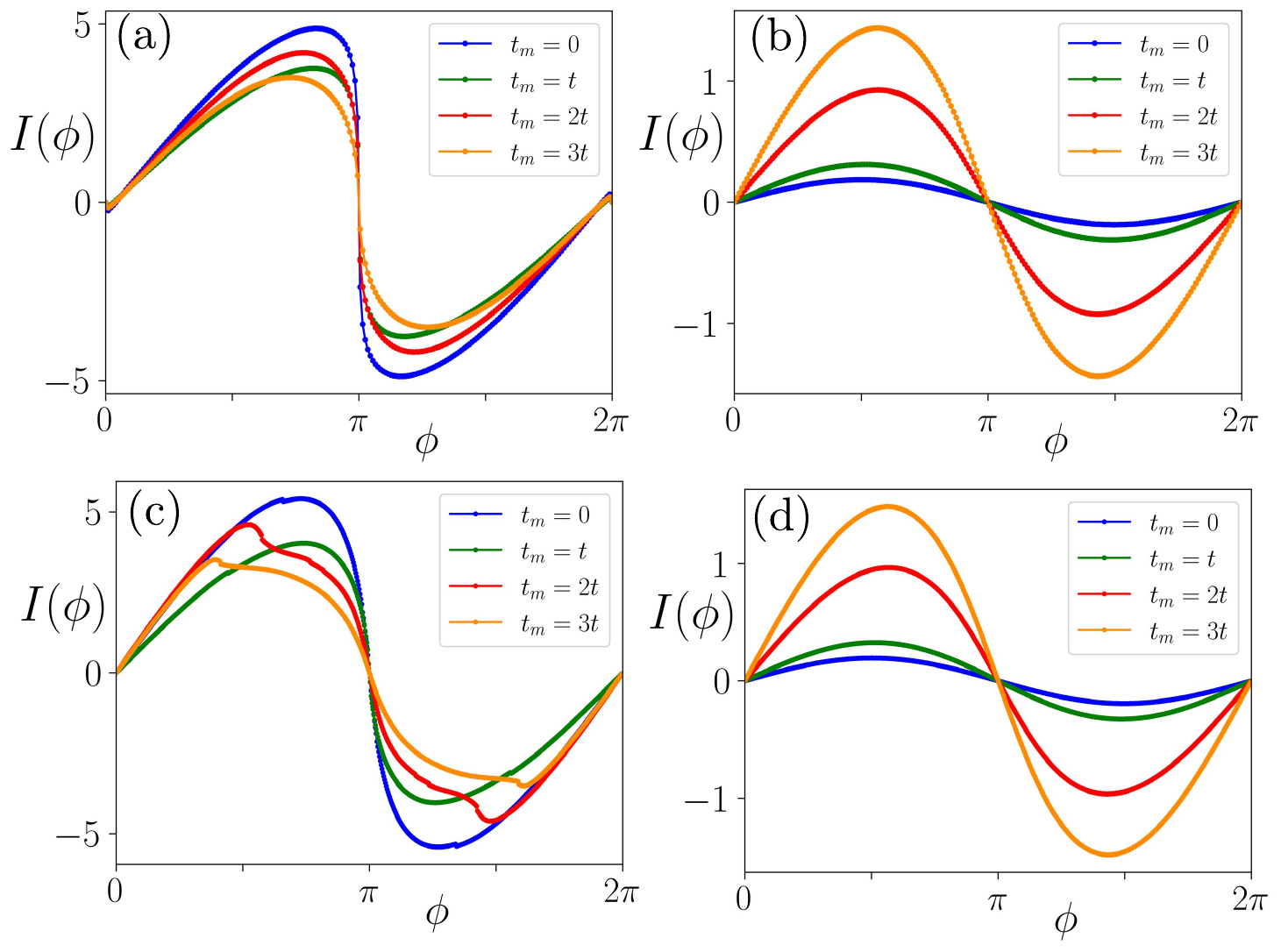}}
	\caption{In panels (a) and (c) we present the behavior of the JC as a function of barrier strength (\(t_m\)) for \(\mu = 2t\), corresponding to the regime hosting MEMs and hybridized modes, in case of \(p_x + p_y\) and \(p_x + ip_y\) SCs, respectively. Conversely, panels (b) and (d) illustrate the variation of JC in the phase dominated by only MZMs (\(\mu = 4.2t\)) for \(p_x + p_y\) and \(p_x + ip_y\) type SCs, respectively. The model parameters used in these calculations are \(\Delta_{0} = 0.5t\), \(t_{0} = t\), and \(t = 1\). We choose a system size of $50\times50$ square lattice.}
	\label{Fig8}
\end{figure}

Now, we study the effect of magnetic barrier on the JC for each of the topological phases. For $\mu<4t$, as the amplitude of spin-dependent hopping \(t_m\) of the magnetic barrier increases, the system makes transition from a phase hosting only MEMs to hybrid topological phase. We depict the variation of $I(\phi)$ as a function of $\phi$ in Fig.~\ref{Fig8}(a) and Fig.~\ref{Fig8}(c) for $p_x + p_y$ and \(p_x + ip_y\) SCs, respectively for various strengths of $t_m$ and fixing $\mu=2t$, $\xi_m=5$. In Fig.~\ref{Fig8}(a), when the system hosts only MEMs i.e., $t_m = 0$ and $t_m=t$, the maximum value of the JC occurs near $\phi = \pi$ and we observe a jump in JC near $\phi=\pi$ which is attributed to the change of parity of the superconducting ground state of $p$-wave SCs~\cite{Alicea_2012} . However, for \(t_m = 2t\) and \(t_m = 3t\), corresponding to the hybrid topological phase, the peak shifts away from \(\phi = \pi\). While the jump remains evident when only MEMs present in the system, its prominence diminishes as the barrier strength increases and the system transits towards the hybrid phase.

In Fig.~\ref{Fig8}(c), a similar behavior is observed for \(t_m = 0\) and \(t_m = t\). However, as \(t_m\) increases further, the system undergoes a phase transition from MEMs to the hybrid phase with coexisting MEMs and MZMs. In this regime, a hump in the JC emerges, which is absent in the \(p_x + p_y\) case due to the gapless nature of the bulk in that system. This hump is a signature of the mixed phase for $p_x+ip_y$ SC as in this case the bulk is gapped and JC carries the signature of mixed phase with both types of Majorana modes.

For $\mu>4t$, the system only hosts MZMs localized at the ends of the magnetic barrier. In the parameter regime \(\mu > 4t\), as shown in Fig.~\ref{Fig8}(b) and Fig.~\ref{Fig8}(d) corresponding to 
\(p_x + p_y\) and \(p_x + ip_y\) SCs, respectively, the MZMs do not exhibit a prominent contribution to the JC as we observe a $2\pi$-periodic Josephson effect. Here, the current primarily flows through the bulk and mostly Cooper pairs contribute to the JC as the MEMs remain gapped out. As the tunneling barrier strength (\(t_m\)) increases, translating from \(t_m = 0\) to \(t_m = 3t\), the contribution of the MZMs to the JC becomes more significant, leading to an overall increase in the JC. This behavior is consistent for both \(p_x + p_y\) and \(p_x + ip_y\) type of pairing symmetries.

In the case of the $p_x + p_y$-wave superconductor, we observe a $4\pi$-periodic Josephson effect at $t_m = 0$, which diminishes as $t_m$ increases. This behavior is consistent with the 1D scenario involving two $p$-wave Kitaev chains coupled via a magnetic weak link as illustrated in 
Appendix~\ref{Appendix-C}. The jump at \( \phi = \pi \) and $t_m = 0$ becomes less sharp due to the finite-size effects in 2D. 
In contrast, for the $p_x + i p_y$-wave superconductor, we do not observe a $4\pi$ periodicity even at $t_m = 0$. 
The reason can be attributed to the fact that in a Josephson junction between two 2D chiral superconductors of the same chirality, both the energy spectrum and the JC exhibit a $2\pi$ periodicity~\cite{Kwon2004}. Furthermore, in both $p_x + p_y$ 
and $p_x + i p_y$ type superconductors, higher harmonics appear in the current-phase relation (CPR). 
This behavior aligns with the expectations as spin-dependent tunneling, which effectively introduces a Zeeman field along the $z$-direction 
(see Appendix~\ref{Appendix-A}). The presence of such effective magnetic field 
at the junction can lead to 
deviations from the conventional sinusoidal CPR and gives rise to higher harmonic contributions in the JC~\cite{Willsch2024, Fukaya2022,Kleiner_2007}.
\section{Stability of MZMs and Josephson Current}\label{Sec:VI}
In this section, we investigate the stability of MZMs in a \( p_x + i p_y \)-wave SC by introducing a localized perturbation at the end points of the magnetic barrier. This perturbation modifies the tunneling Hamiltonian \( H_{LR} \), as defined in Eq.~(\ref{Eq. Coupling_Ham_H_LR}), and is applied specifically at the barrier terminations located at $y = 0 \text{ and } y = L_y, \text{ with } x = L_x/2$.

The corresponding perturbative Hamiltonian is given by:
\begin{equation}
	\mathcal{H}_{p} = \sum_{y,\sigma} c^\dagger_{L_x/2,y,\sigma} \tilde{t}_1 \, c_{L_x/2+a,y,\sigma} + \text{h.c.}\ ,
	\label{Eq. H_perturb}
\end{equation}
where \( \tilde{t}_1 \) denotes the strength of the perturbation.

\begin{figure}[h]
	\centering
	\includegraphics[scale=0.48]{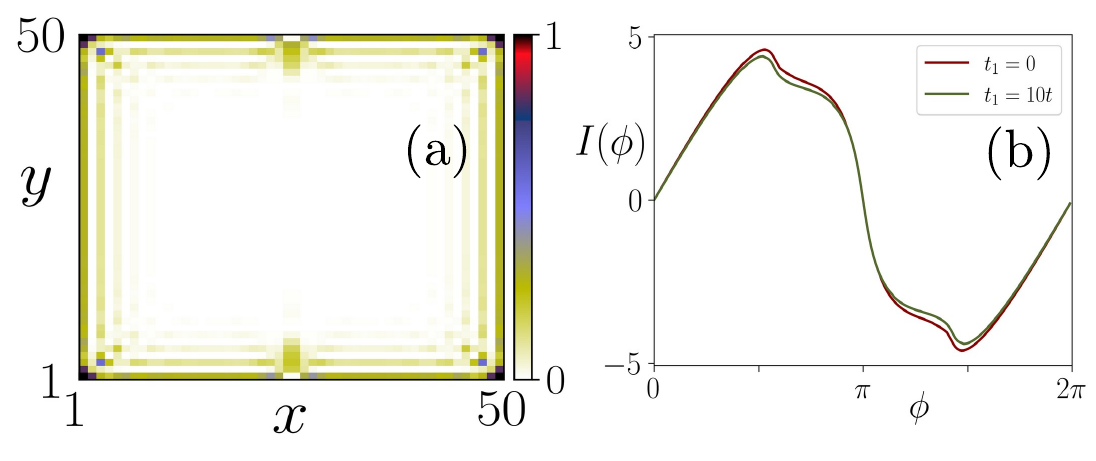}
	\caption{(a) In panel (a), we showcase the persistence of MZMs in LDOS spectra under a perturbation of strength \( \tilde{t}_1 = 10t \). 
	In panel (b), we show the JC for two cases: \( \tilde{t}_1 = 0 \) (red curve) and \( \tilde{t}_1 = 10t \) (green curve). 
    We use the other model parameters as \( \mu = 2t \), \( t_m = 2t \), \( t_0 = t \), \( \Delta_0 = 0.5t \), \( \xi_m = 5t \), and \( t = 1 \), on a finite size system of \( 50 \times 50 \) lattice sites. }
	\label{fig:Fig9}
\end{figure}

We consider a strong perturbation with \( \tilde{t}_1 = 10t \). Our numerical results exhibit that the MZMs remain robust in the presence of such local perturbation. As shown in Fig.~\ref{fig:Fig9}(a), the MZMs are not destroyed. However, they are slightly displaced from the ends of the magnetic barrier to adjacent lattice sites along the junction. Nevertheless, the characteristic zero-energy feature associated with MZMs continues to appear in the LDOS, indicating their persistence. In Fig.~\ref{fig:Fig9}(b), we depict the JC for both the unperturbed 
(\( \tilde{t}_1 = 0 \)) and perturbed (\( \tilde{t}_1 = 10t \)) cases. The CPR remains qualitatively unchanged, further confirming the robustness of the hump signature indicating the mixed phase (presence of both MEMs and MZMs) for \( p_x + i p_y \) SC.


\section{Summary and discussion}\label{Sec:VII}
To summarize, in this article, we investigate the effects due to a spatially varying magnetic barrier acting as a junction between two $p$-wave SCs, considering both types of pairing symmetries: $p_x + p_y$ and $p_x + i p_y$. Our analysis reveals a rich topological phase structure, with the magnetic barrier extending the topological superconducting region in the $p$-wave phase that inherently hosts MEMs localized at the system's boundaries. We identify the emergence of three distinct topological regions as a function of the magnetic barrier strength: (i) MEMs localized on the edges of the 2D $p$-wave SC, (ii) MZMs localized at the ends of the magnetic chain, and (iii) hybridized regime, where MEMs and MZMs coexist. Specifically, in the region where MZM exists, we demonstrate the ability to tune the number of zero-energy modes. This can be carried way by varying both the strength and the periodicity of the spatially modulated barrier, providing a pathway for precise control of Majorana modes.

The identification of these distinct topological phases is further analyzed by characterizing the LDOS under varying phase differences across the junction. Furthermore, we compute the JC across the junction to examine the behavior of different Majorana modes, which provide insights into the interplay between pairing symmetries, chemical potential, magnetic barrier strength (spin-dependent), and topological phase transitions. More specifically, we distinguish between MEMs and MZMs through JC signatures. For $p_x+ip_y$-type pairing, the hybrid topologial phase exhibits a pronounced hump which could be an interesting observation from an experimental perspective. 

In recent times, several experiments have reported the coexistence of non-collinear magnetic textures and superconductivity~\cite{Wiesendanger2021,Beck2021,Wang2021,Schneider2022,Richard2022,Wiesendanger2022}. These experiments suggest that such coexistence is possible when the characteristic length of the magnetic spin texture is comparable to the superconducting coherence length. Therefore, this condition should be a necessary criterion for constructing a JJ where the magnetic spin texture acts as a tunnel barrier. On the other hand, there are very few materials reported in the literature that exhibit $p$-wave superconductivity. One of the potential candidates is $\rm{Sr}_2\rm{RuO}_4$; however, its $p$-wave nature remains debatable~\cite{Pustogow2019,Aaron2021}. Intensive research is still ongoing, and other potential candidates include uranium-based heavy-fermion compounds, among which $\rm{UTe}_2$ appears to be the most promising~\cite{Jiao2020}. Possible materials for a magnetic textured tunnel barrier include Mn, Cr, etc~\cite{Wiesendanger2022,Chatterjee2023_PRBL}. Moreover, 
we believe that our study can provide a viable platform to distinguish between MEMs and MZMs through JC signatures across magnetic texture coupled JJ.
%
%

\subsection*{Acknowledgments}
We acknowledge Department of Atomic Energy (DAE), Govt. of India for providing the financial support. We also acknowledge SAMKHYA: High-Performance Computing Facility provided by Institute of Physics, Bhubaneswar and the two workstations provided by the Institute of Physics, Bhubaneswar from the DAE APEX project for numerical computations.

\subsection*{Data Availibility Statement}
The datasets generated and analyzed during the current study are available from the corresponding author upon reasonable request.

\appendix
{\section{Derivation of the Effective 1D Hamiltonian}\label{Appendix-A}}
In this appendix, we derive an effective 1D Hamiltonian starting from Eq.~(\ref{Eq.BdG_Ham}). To simplify the problem, we consider a system with \( L_x = 2 \) and \( L_y \rightarrow \infty \), applying periodic boundary condition (PBC) along the \( y \)-direction to approximate the bulk limit. This setup reduces the system to two coupled superconducting chains aligned along the \( y \)-axis with magnetic weak link between them. We aim to develop a simplified 1D model that can be solved exactly to provide analytical insights into our main results.

The system consists of left ($L$) and right ($R$) superconducting chains coupled via a harmonic spin-textured barrier \( V_{LR}(y) \). The $L$ and $R$ regions are characterized by superconducting phases \( \phi_L = -\phi/2 \) and \( \phi_R = +\phi/2 \), respectively. The total Hamiltonian can then be expressed as~\cite{Sardinero}

\begin{equation}
	H =
	\begin{pmatrix}
		H_L & V_{LR}(y) \\
		V_{LR}^\dagger(y) & H_R
	\end{pmatrix}\ ,
\end{equation}
where, the left and right Hamiltonians \( H_{L,R} \) are given by
\begin{equation*}
    \begin{aligned}
	H_{L,R} = &-[\mu - 2t\cos k_y]\, \tau_z \sigma_0 + \Delta \sin k_y [ \cos \phi_{L,R} \tau_x \sigma_y \\
	&- \sin \phi_{L,R}  \tau_y \sigma_y ]\ ,
    \end{aligned}
\end{equation*}
and the inter-chain coupling reads
\begin{equation*}
V_{LR} = -t_0 \tau_z \sigma_0 - t_m \left( \sin \theta_y \, \tau_0 \sigma_x + \cos \theta_y \, \tau_0 \sigma_z \right)\ .
\end{equation*}

Combining these, the full Hamiltonian becomes
\begin{equation}
	\begin{aligned}
		H &= -[\mu + 2t\cos k_y]\, u_0 \tau_z \sigma_0 + \Delta \sin k_y [ \cos\frac{\phi}{2} \, u_0 \tau_x \sigma_y\\ &+\sin\frac{\phi}{2} \, u_z \tau_y \sigma_y ] 
		- t_0 u_x \tau_z \sigma_0 - t_m (\sin \theta_y \, u_x \tau_0 \sigma_x \\
		&+ \cos \theta_y \, u_1 \tau_0 \sigma_3 )\ ,
	\end{aligned}
\end{equation}
where $u$ represents Pauli matrices acting on the left and right subspace.

To remove the \( y \)-dependence from the angle \( \theta_y \), we perform a local unitary transformation using the operator

\begin{equation}
	U = u_0 \left[ \cos \frac{\theta_y}{2} \, \tau_z \sigma_0 - i \sin \frac{\theta_y}{2} \, \tau_z \sigma_y \right]\ .
\end{equation}

The effective Hamiltonian in the rotated frame is then given by \( H_{\text{eff}} = U^\dagger H U \). For the low-energy, we approximate:
\[
\cos k_y \approx 1 - \partial_y^2, \quad \sin k_y \approx -i \partial_y\ ,
\]
Substituting these into the transformed Hamiltonian yields
\begin{equation}
	\begin{aligned}
		H_{\text{eff}} &= -\left[ \mu_{\text{eff}}+ 2t\cos k_y  \right] u_0 \tau_z \sigma_0 - \frac{2\pi t}{\xi_m} \sin k_y \, u_0 \tau_z \sigma_y \\
		&\quad + \Delta \cos\frac{\phi}{2} \left( \sin k_y u_0 \tau_x\sigma_y + \frac{\pi}{\xi_m} u_0\tau_x\sigma_0 \right)- t_0 u_x \tau_z \sigma_0  \\
		&\quad  - t_m u_x \tau_0 \sigma_z +\Delta \sin\frac{\phi}{2} \left( \sin k_y u_z \tau_y\sigma_y + \frac{\pi}{\xi_m} u_z\tau_y\sigma_0 \right)\ .
	\end{aligned}
\label{1DEffHam}
\end{equation}

Note that, the effective chemical potential is given by $\mu_{\text{eff}} = \mu - \frac{\pi^2 t}{\xi_m^2}$, indicating a reduction by the term \(\frac{\pi^2 t}{\xi_m^2}\). Effective spin-orbit coupling emerges in the Hamiltonian, $H_{\text{eff}}$ through the term \(\frac{2\pi t}{\xi_m} \sin k_y\). Additionally, the superconducting pairing is modified by an effective factor of \(\frac{\pi}{\xi_m}\). An effective Zeeman term 
also arises from spin dependent hopping strength, \(t_m\), along the \(z\) direction.

\vspace {0.5cm}
\section{Topological Characterization}{\label{Appendix-B}}
In this appendix, we present the topological invariants associated with MEMs and MZMs, characterized respectively by the Chern number and the winding number.

\subsection{Chern Number}\label{Appendix-B1}
\begin{figure}
	\centering
	\includegraphics[scale=0.4]{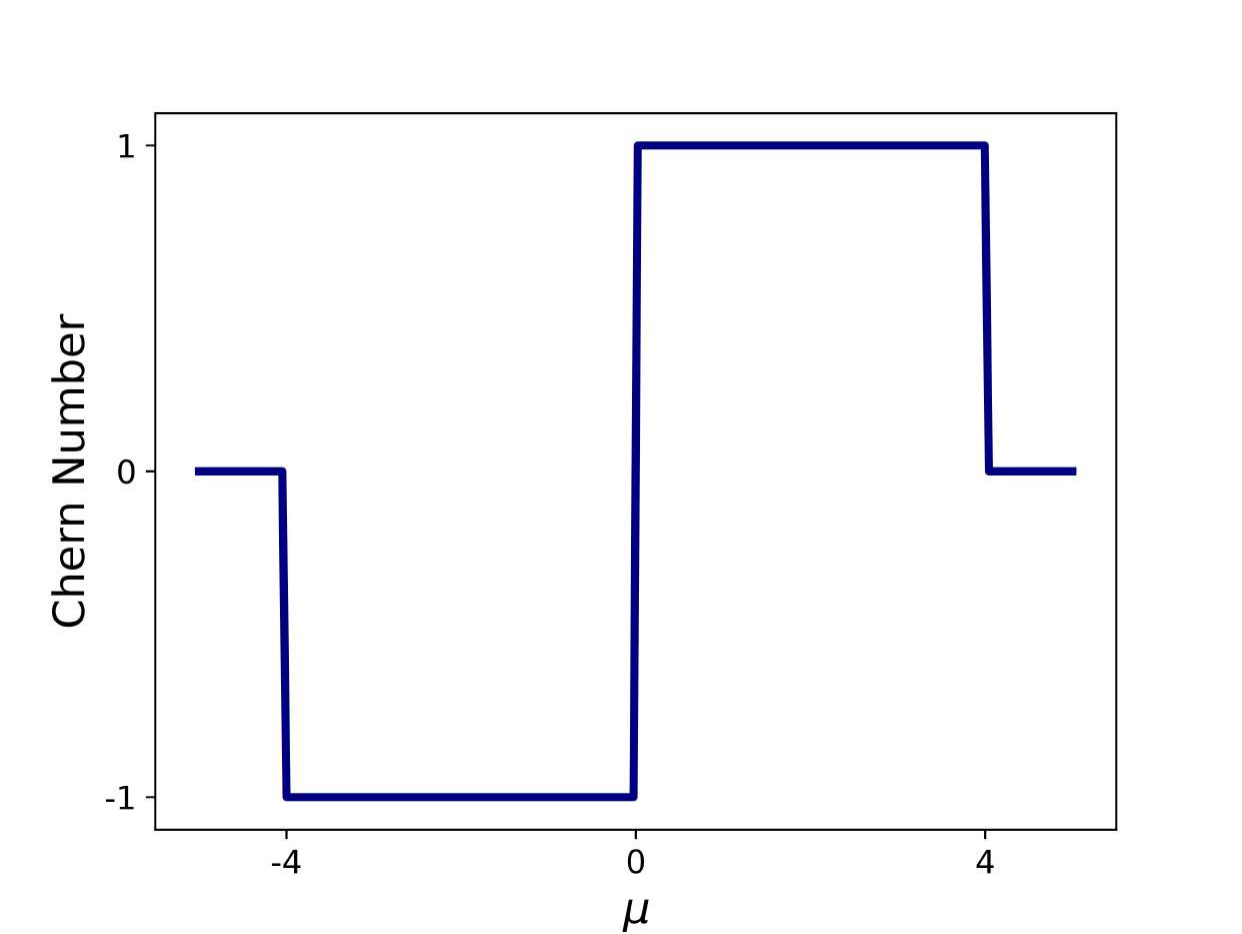}
	\caption{Chern number is depicted as a function of chemical potential $\mu$ (along the range of $\mu$ mentioned in Fig.~\ref{Fig3}(a) 
	of the main text) for the $p_x + ip_y$-wave SC when $t_{m}=0$.}
	\label{fig:Fig10}
\end{figure}

We compute the Chern number using the Fukui method~\cite{Fukui_2005} in the absence of spin-dependent tunneling ($t_m = 0$), where only the normal tunneling amplitude ($t_0$) is present. In this regime, the system behaves as a single 2D $p$-wave superconductor hosting MEMs localized at its boundary. For the $p_x + ip_y$-type pairing, the Chern number is calculated and shown in Fig.~\ref{fig:Fig10}. In contrast, for the 
$p_x + p_y$ configuration, the bulk spectrum becomes gapless, rendering the Chern number ill-defined.

For the $p_x + ip_y$ SC, the Chern number takes the value,
\begin{itemize}
	\item $C = 0$ in the trivial regime,
	\item $C = +1$ for $0 < \mu < 4t$, corresponding to the topological regime hosting a single chiral MEM, and
	\item $C = -1$ in another topological phase depending on the chemical potential,
\end{itemize}
as illustrated in Fig.~\ref{fig:Fig10}.

\subsection{Winding Number}\label{Appendix-B2}
To characterize the MZMs, we analyze the effective 1D model (Eq.~(\ref{1DEffHam})) and compute the winding number. For simplicity, we set the superconducting phase $\phi = 0$. We define a chiral symmetry operator $C = u_0 \tau_y \sigma_y$, under which the Hamiltonian can be recast
into an off-diagonal form via a unitary transformation $U_c$, composed of the eigenvectors of $C$. In the chiral basis, the Hamiltonian takes 
the form
\begin{equation}
	H =
	\begin{pmatrix}
		0 & H^+(k) \\
		H^-(k) & 0
	\end{pmatrix}\ ,
\end{equation}
where, $H^{\pm}(k)$ denotes the $4 \times 4$ matrices defined within the chiral subspace.

The winding number $W$, a topological invariant, is then calculated as~\cite{Chiu2016RMP,Mondal2023b}:
\begin{equation}
	W = \left| \pm \frac{i}{2\pi} \int_{-\pi}^{\pi} dk \, \text{Tr} \left\{ \left[ H^\pm(k) \right]^{-1} \partial_k H^\pm(k) \right\} \right|\ .
\end{equation}

\begin{figure}
	\centering
	\includegraphics[scale=0.35]{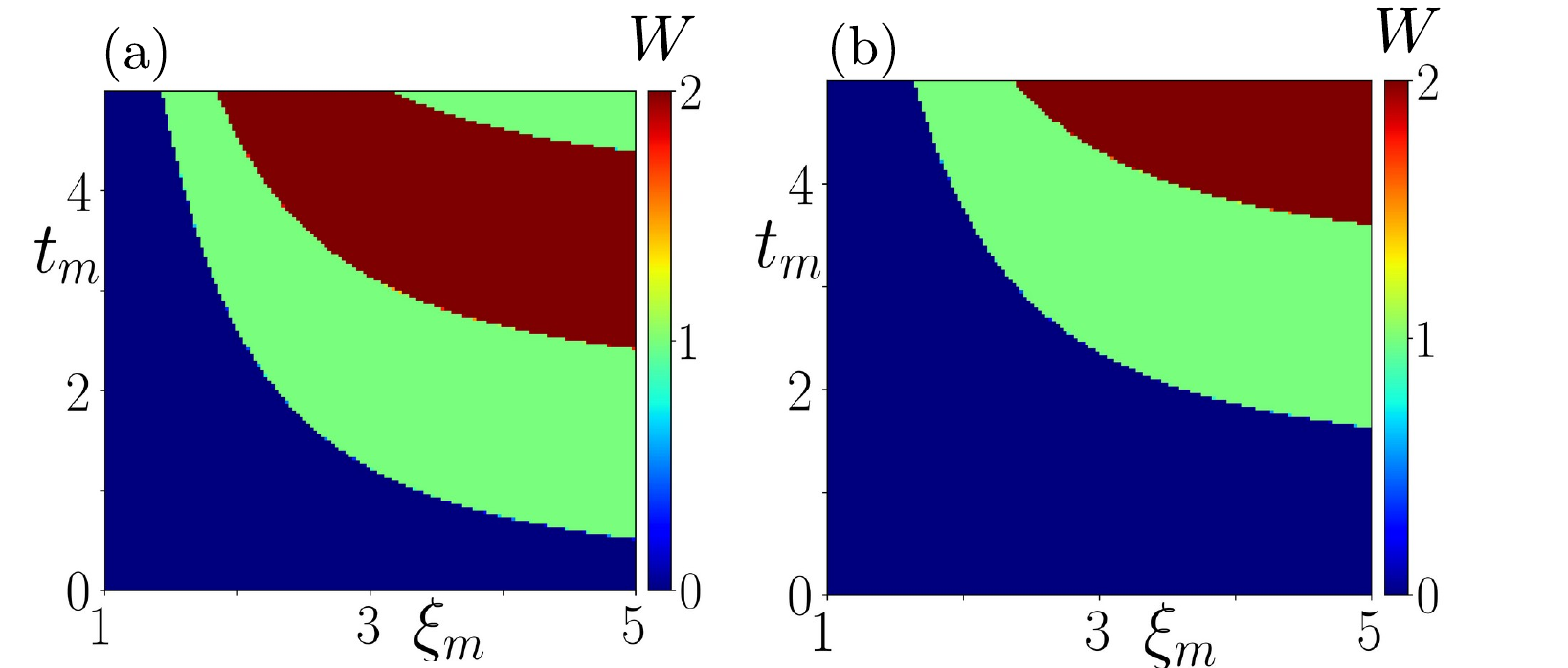}
	\caption{We illustrate the variation of winding number $W$ in $t_m - \xi_m$ plane for the effective 1D model choosing \(\mu = 3t\) in panel (a) and in panel (b) considering \(\mu = 4.2t\). The other model parameters are considered to be $\Delta=0.5t$, $t_0=t$ and $t=1$.}
	\label{fig:Fig11}
\end{figure}

We evaluate $W$ in the plane of the spin-dependent tunneling $t_m$ and the spatial modulation vector $\xi_m$, for two values of the chemical potential: $\mu = 3t$ and $\mu = 4.2t$ (see Figs.~\ref{fig:Fig11}(a) and (b) respectively). These correspond to hybrid and fully topological phases in the full 2D lattice model, respectively. The computed winding number takes values $W = 1$ or $2$, indicating the presence of one or 
two MZMs at each end of the effective 1D junction along $y$ direction.

\section{Magnetic weak link between two 1D \(p\)-wave SCs}{\label{Appendix-C}}

\begin{figure}[h]
	\centering
	\includegraphics[scale=0.5]{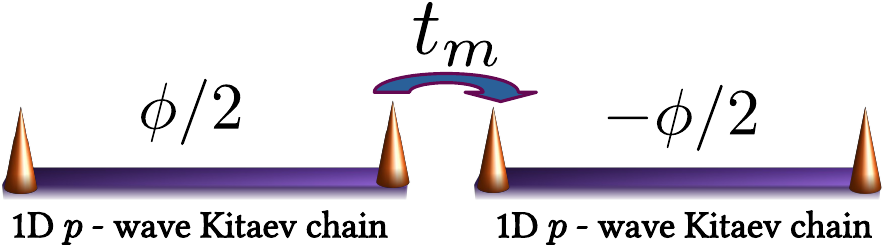}
	\caption{We schematically depict the Josephson junction comprised of two 1D \(p\)-wave Kitaev chains with superconducting phase difference 
	of $\phi$, connected via a magnetic weak link with strength \(t_m\).}
	\label{fig:Fig12}
\end{figure}

In this appendix, we numerically study the behavior of a Josephson junction comprised of a magnetic weak link connecting two 1D \(p\)-wave Kitaev chains as shown in Fig.~\ref{fig:Fig12}. Our focus is on how varying the magnetic coupling strength, denoted by \(t_m\), influences the energy spectrum and the corresponding JC.

\begin{figure}
	\centering
	\includegraphics[scale=0.31]{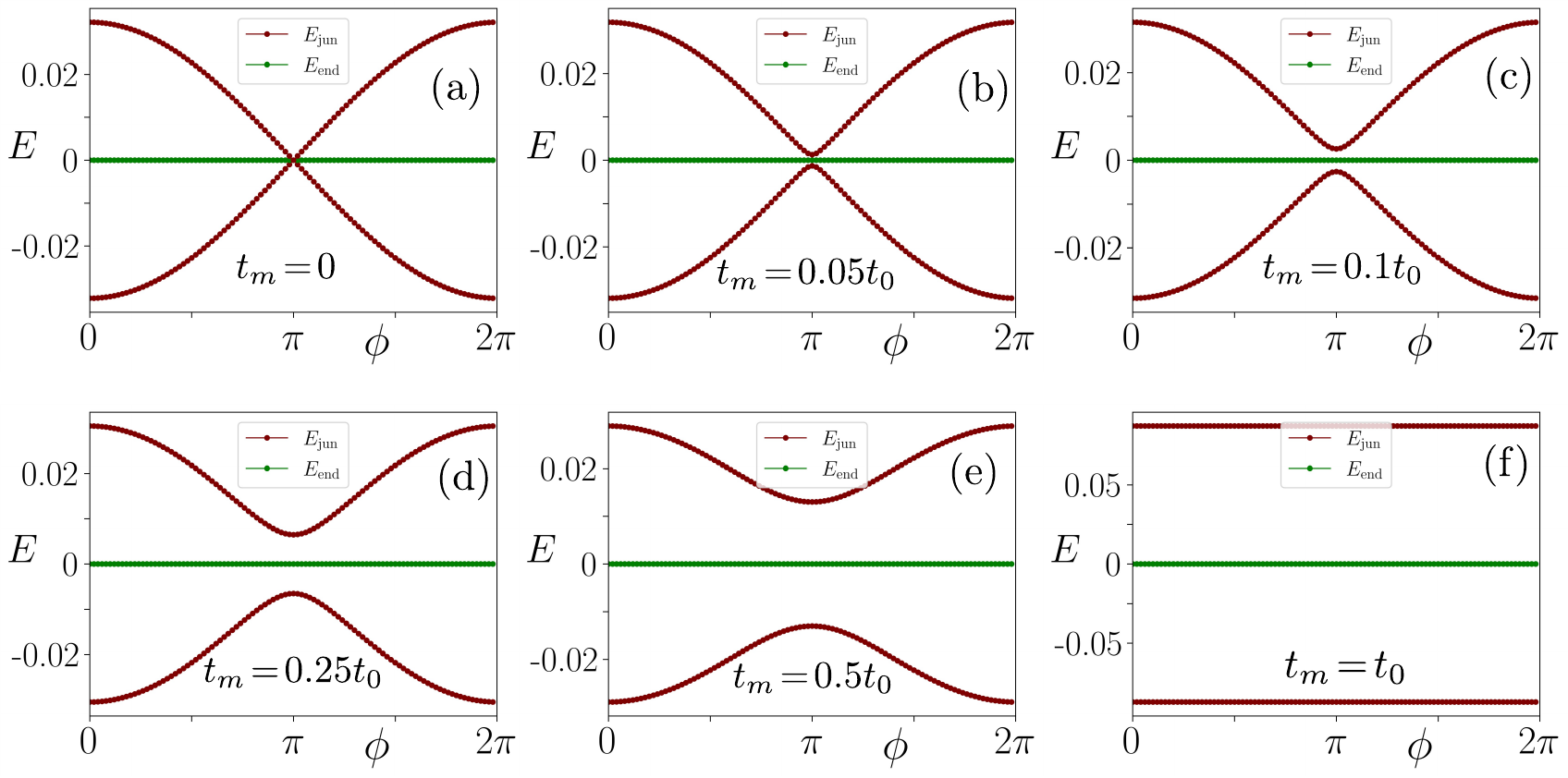}
	\caption{We display the energy-phase (\(E\)-\(\phi\)) relation displaying the evolution of energy levels with phase difference $\phi$. Panel (a) \(t_m = 0\): protected zero-energy crossing at \(\phi = \pi\) is depicted. In panels (b), (c), (d) and (e) 
	we consider \(t_m \neq 0\): gap opens, lifting the degeneracy. In panel (f), we set $t_m=t_0$ and there is no phase effect.}
	\label{fig:Fig13}
\end{figure}

Initially, when \(t_m = 0\), the system hosts localized MZMs at the junction (with energy $E_{\rm jun}$) and outer ends (with energy $E_{\rm end}$). When the phase difference becomes $\phi=\pi$, two branches of $E_{\rm jun}$ crosses each other, signaling the change in fermionic parity of the ground state resulting in the characteristic \(4\pi\)-periodic energy-phase relation as illustrated in Fig.~\ref{fig:Fig13}(a).


\begin{figure}
	\includegraphics[scale=0.37]{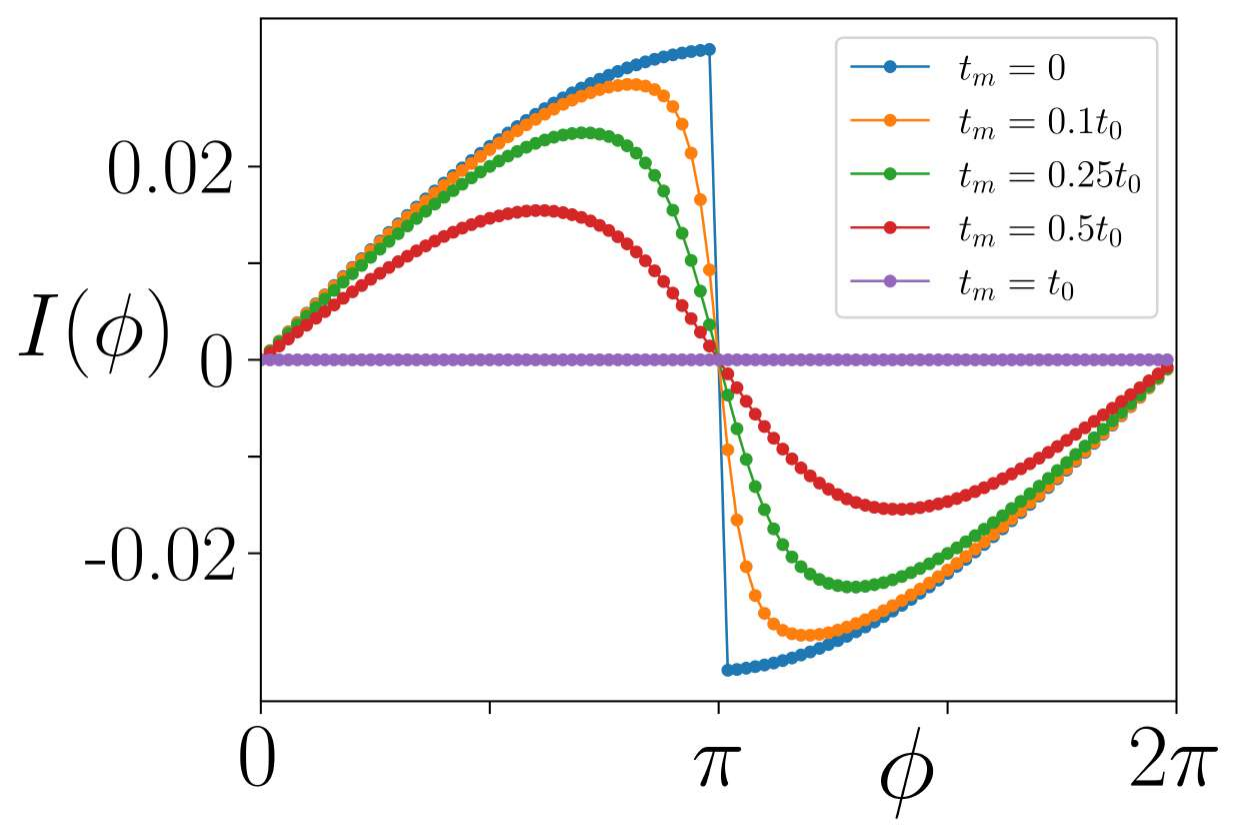}
	\caption{We showcase the JC ($I(\phi)$) as a function of phase bias ($\phi$) considering the 1D $p$-wave JJ setup in presence of a magnetic link with various strengths $t_m$.}
	\label{fig:Fig14}
\end{figure}


When the magnetic coupling \(t_m\) is turned on, the zero-energy crossing at \(\phi = \pi\) disappears as illustrated in 
Figs.~\ref{fig:Fig13}(b)-(f). This indicates the hybridization of the MZMs localized at the junction, lifting the degeneracy and resulting in a gap opening at $\phi=\pi$, thereby destroying the \(4\pi\)-periodicity in the energy spectrum.


This transition is also reflected in the behavior of JC. As shown in Fig.~\ref{fig:Fig14}, when \(t_m = 0\), the current manifests 
a \(4\pi\)-periodic component (discontinous jump) due to the presence of topologically protected MZMs. However, even a small $t_{m}\ll t_{0}$ suppresses this behavior, restoring the conventional \(2\pi\)-periodicity.
As the magnetic coupling approaches the normal tunneling strength (\(t_m \approx t_0\)), the JC eventually becomes vanishingly small (see Fig.~\ref{fig:Fig14}).

\bibliographystyle{apsrev4-2mod}
\bibliography{bibfile}

\end{document}